\documentclass[twocolumn,trackchanges]{aastex631}

\usepackage{xspace,upgreek,booktabs}
\newcommand\um{\ensuremath{\upmu\mathrm{m}}\xspace}

\begin{document}

\title{PHANGS-JWST First Results: The 21 \um Compact Source Population}
\turnoffeditone
\turnoffedittwo
\author[0000-0002-8806-6308]{Hamid Hassani}
\affiliation{Department of Physics, University of Alberta, Edmonton, Alberta, T6G 2E1, Canada}

\author[0000-0002-5204-2259]{Erik Rosolowsky}
\affiliation{Department of Physics, University of Alberta, Edmonton, Alberta, T6G 2E1, Canada}

\author[0000-0002-2545-1700]{Adam K. Leroy}
\affiliation{Department of Astronomy, The Ohio State University, 140 West 18th Avenue, Columbus, Ohio 43210, USA}

\author[0000-0003-0946-6176]{Médéric~Boquien}
\affiliation{Centro de Astronomía (CITEVA), Universidad de Antofagasta, Avenida Angamos 601, Antofagasta, Chile}

\author[0000-0003-0946-6176]{Janice C. Lee}
\affiliation{Gemini Observatory/NSF’s NOIRLab, 950 N. Cherry Avenue, Tucson, AZ, USA}
\affiliation{Steward Observatory, University of Arizona, 933 N Cherry Ave,Tucson, AZ 85721, USA}

\author[0000-0003-0410-4504]{Ashley.~T.~Barnes}
\affiliation{Argelander-Institut f\"{u}r Astronomie, Universit\"{a}t Bonn, Auf dem H\"{u}gel 71, 53121, Bonn, Germany}

\author[0000-0002-2545-5752]{Francesco Belfiore}
\affiliation{INAF — Arcetri Astrophysical Observatory, Largo E. Fermi 5, I-50125, Florence, Italy}

\author[0000-0003-0166-9745]{F. Bigiel}
\affiliation{Argelander-Institut f\"ur Astronomie, Universit\"at Bonn, Auf dem H\"ugel 71, 53121 Bonn, Germany}

\author[0000-0001-5301-1326]{Yixian Cao}
\affiliation{Max-Planck-Institut f\"ur Extraterrestrische Physik (MPE), Giessenbachstr. 1, D-85748 Garching, Germany}

\author[0000-0002-5635-5180]{M\'elanie Chevance}
\affiliation{Universit\"{a}t Heidelberg, Zentrum f\"{u}r Astronomie, Institut f\"{u}r Theoretische Astrophysik, Albert-Ueberle-Stra{\ss}e 2, D-69120 Heidelberg, Germany}
\affiliation{Cosmic Origins Of Life (COOL) Research DAO, coolresearch.io}

\author[0000-0002-5782-9093]{Daniel~A.~Dale}
\affiliation{Department of Physics and Astronomy, University of Wyoming, Laramie, WY 82071, USA}

\author[0000-0002-4755-118X]{Oleg V. Egorov}
\affiliation{Astronomisches Rechen-Institut, Zentrum f\"{u}r Astronomie der Universit\"{a}t Heidelberg, M\"{o}nchhofstra\ss e 12-14, D-69120 Heidelberg, Germany}

\author[0000-0002-6155-7166]{Eric Emsellem}
\affiliation{European Southern Observatory, Karl-Schwarzschild-Stra{\ss}e 2, 85748 Garching, Germany}
\affiliation{Univ Lyon, Univ Lyon1, ENS de Lyon, CNRS, Centre de Recherche Astrophysique de Lyon UMR5574, F-69230 Saint-Genis-Laval France}

\author[0000-0001-5310-467X]{Christopher M. Faesi}
\affiliation{University of Connecticut, Department of Physics, 196A  Auditorium Road, Unit 3046, Storrs, CT, 06269}

\author[0000-0002-3247-5321]{Kathryn~Grasha}
\affiliation{Research School of Astronomy and Astrophysics, Australian National University, Canberra, ACT 2611, Australia}   
\affiliation{ARC Centre of Excellence for All Sky Astrophysics in 3 Dimensions (ASTRO 3D), Australia}

\author[0000-0002-0432-6847]{Jaeyeon Kim}
\affiliation{Universit\"{a}t Heidelberg, Zentrum f\"{u}r Astronomie, Institut f\"{u}r Theoretische Astrophysik, Albert-Ueberle-Stra{\ss}e 2, D-69120 Heidelberg, Germany}

\author[0000-0002-0560-3172]{Ralf S.\ Klessen}
\affiliation{Universit\"{a}t Heidelberg, Zentrum f\"{u}r Astronomie, Institut f\"{u}r Theoretische Astrophysik, Albert-Ueberle-Stra{\ss}e 2, D-69120 Heidelberg, Germany}
\affiliation{Universit\"{a}t Heidelberg, Interdisziplin\"{a}res Zentrum f\"{u}r Wissenschaftliches Rechnen, Im Neuenheimer Feld 205, D-69120 Heidelberg, Germany}

\author[0000-0001-6551-3091]{Kathryn Kreckel}
\affiliation{Astronomisches Rechen-Institut, Zentrum f\"{u}r Astronomie der Universit\"{a}t Heidelberg, M\"{o}nchhofstra\ss e 12-14, 69120 Heidelberg, Germany}

\author[0000-0002-8804-0212]{J.~M.~Diederik~Kruijssen}
\affiliation{Cosmic Origins Of Life (COOL) Research DAO, coolresearch.io}

\author[0000-0003-3917-6460]{Kirsten L. Larson}
\affiliation{AURA for the European Space Agency (ESA), Space Telescope Science Institute, 3700 San Martin Drive, Baltimore, MD 21218, USA}

\author[0000-0002-6118-4048]{Sharon E. Meidt}
\affiliation{Sterrenkundig Observatorium, Universiteit Gent, Krijgslaan 281 S9, B-9000 Gent, Belgium}

\author[0000-0002-4378-8534]{Karin M. Sandstrom}
\affiliation{Department of Physics, University of California, San Diego, 9500 Gilman Drive, San Diego, CA 92093, USA}

\author[0000-0002-3933-7677]{Eva Schinnerer}
\affiliation{Max-Planck-Institut f\"ur Astronomie, K\"onigstuhl 17, D-69117 Heidelberg, Germany}

\author[0000-0002-8528-7340]{David A. Thilker}
\affiliation{Department of Physics and Astronomy, The Johns Hopkins University, Baltimore, MD 21218, USA}

\author[0000-0002-7365-5791]{Elizabeth~J.~Watkins}
\affiliation{Astronomisches Rechen-Institut, Zentrum f\"{u}r Astronomie der Universit\"{a}t Heidelberg, M\"{o}nchhofstra\ss e 12-14, 69120 Heidelberg, Germany}

\newcommand{\STScI}{\affiliation{Space Telescope Science Institute, 3700 San Martin Drive, Baltimore, MD 21218, USA}}
\author[0000-0002-3784-7032]{Bradley~C.~Whitmore}
\STScI

\author[0000-0002-0012-2142]{Thomas G. Williams}
\affiliation{Sub-department of Astrophysics, Department of Physics, University of Oxford, Keble Road, Oxford OX1 3RH, UK}
\affiliation{Max-Planck-Institut f\"ur Astronomie, K\"onigstuhl 17, D-69117 Heidelberg, Germany}

\suppressAffiliations



\begin{abstract}

We use PHANGS-JWST data to identify and classify 1271 compact 21\,\um sources in four nearby galaxies using MIRI F2100W data. We identify sources using a dendrogram-based algorithm, and we measure the background-subtracted flux densities for JWST bands from 2\,$\um$ to 21\,$\um$. Using the SED in JWST as well as HST bands, plus ALMA and MUSE/VLT observations, we classify the sources by eye. Then we use this classification to define regions in color-color space, and so establish a quantitative framework for classifying sources. We identify 1085 sources as belonging to the ISM of the target galaxies with the remainder being dusty stars or background galaxies.  These 21 \um sources are strongly spatially associated with \ion{H}{2} regions ($>92\%$ of sources), while 74\% of sources are coincident with a stellar association defined in the HST data. Using SED fitting, we find that the stellar masses of the 21 \um sources span a range of 10$^{2}$ to 10$^{4}~M_\odot$ with mass-weighted ages down to 2 Myr. There is a tight correlation between attenuation-corrected H$\alpha$ and 21~\um luminosity for $L_{\nu,\mathrm{F2100W}}>10^{19}~\mathrm{W~Hz}^{-1}$. Young embedded source candidates selected at 21~\um are found below this threshold and have $M_\star < 10^{3}~M_\odot$. 
\end{abstract}



\section{Introduction} \label{sec:intro}

Dust grains and polycyclic aromatic hydrocarbons (PAHs) in the interstellar medium (ISM) play a central role in shaping the radiation field in galaxies, converting short wavelength light from stars and other emitters into long wavelength emission in the infrared \citep[IR][]{galliano18}.  The mid-IR emission observed by JWST is generated by small dust grains, which are heated stochastically to temperatures of $\gtrsim$100--150\,K to emit photons at mid-infrared ($5 \lesssim \lambda/\um < 60$) wavelengths \citep{dl01}. PAHs re-emit absorbed radiation in discrete spectral band features from $3 < \lambda/\um \lesssim 21$, which correspond to the stretching and bending modes of bonds in the large molecules \citep{allamandola89, Li20}. Because the IR radiation is significantly less affected by absorption, dust and PAH emission provides a vital low-extinction view into the densest regions of galaxies. 

Thanks to previous generations of mid-IR observatories, the properties of warm dust and PAHs have been broadly surveyed. Such studies have illustrated the crucial role of IR observations in understanding the process of star formation in galaxies since star formation occurs in high extinction regions \citep[e.g.,][]{SPIT,Jarrett}.  In particular, embedded high mass stars create strong radiation fields that enhance the mid-IR emission so that the mid-IR can be used as a tracer for star formation \citep{calzetti07, kennicutt12}. Mid-IR data are usually paired with a short wavelength tracer of unobscured star formation to approximate the full star formation rate \citep{kennicutt09, hao11, z0mgs}.  However, mid-IR emission is not a linear tracer of the star formation rate at high resolution.  In addition to the decorrelation of different stages of the star formation process at small ($<1~\mathrm{kpc}$) scales \citep[e.g.,][]{schruba10, onodera10, kruijssen14}, the mid-IR emission also arises from the diffuse, average interstellar radiation field heating the neutral ISM \citep{boquien15, SANDSTROM1_PHANGSJWST, LEROY1_PHANGSJWST}.  

Understanding the nature of the correlation between star formation and the mid-IR has been challenging because the resolution of the previous generation of observatories had comparatively poor resolution: $6\farcs5$ for the {\it Spitzer}/MIPS 24 \um band \citep{spitzermission} and $11\farcs 9$ for the WISE 22 \um band \citep{wisemission}.  The study of individual star forming regions at $<100$~pc scales was thus limited to the Milky Way, the Local Group, and the nearest ($d\lesssim 3$~Mpc) galaxies \citep[e.g.,][]{mipsgal, sage, verley07,peeters02,chastenet19}. Whereas PAH emission is distributed throughout the neutral ISM, several authors have noted the spatial correspondence between H$\alpha$ emission and the mid-IR continuum \citep{rice90, helou04, verley07, relano09}.  These moderately resolved extragalactic studies suggested that the PAH emission tends to be found at the edges of \ion{H}{2} regions, which is supported by the high physical resolution observations of individual Milky Way objects \citep[e.g.][]{mipsgal}.  Given the close link between star formation and the IR emission, extragalactic observations have used the mid-IR to identify compact regions of star formation.  Such studies characterize the population of stellar clusters and compact association that are in the process of forming and estimate the properties of young stellar structures \citep{sharma11} or the characteristic spacing between regions \citep{elmegreen19}.  




With the launch of JWST, we now have the opportunity to make sensitive, higher resolution observations at mid-IR wavelengths.  In particular, the JWST/MIRI observations at $5-28~\um$ have an order of magnitude improvement in resolution over {\it Spitzer}/MIPS and superior sensitivity.  This capability offers a new opportunity to push studies of resolved mid-IR emission to galaxies well beyond the Local Group.  

This paper explores the relationship between compact sources seen in MIRI 21 \um observations and other tracers of star formation.  We use JWST observations of the first four galaxies to be observed for the Physics at High Angular resolution in Nearby GalaxieS (PHANGS) Treasury program (02107, PI J. Lee) in the near- and mid-IR \citep{LEE_PHANGSJWST}.  We combine these new observations with the rich set of supporting data about the star formation process gathered as part of the PHANGS surveys.  We aim to understand the nature of the compact 21~\um sources seen in the JWST imaging data, extending the studies from the Local Group to a more diverse set of star-forming environments and determining if these sources include a set of truly embedded star-forming regions.  But even with the $0\farcs67$ FWHM of the 21 \um filter with JWST, these maps are not resolving individual stars or even stellar clusters \citep[e.g.,][]{RODRIGUEZ_PHANGSJWST}.  We thus face a challenge of how to identify and extract a uniform set of regions.

To achieve these goals, we first develop a method to find compact sources in images with large amounts of diffuse emission (Section \ref{sec:source}).  Then, we use the colors (flux ratios) of the sources in JWST bands to reject objects that are unlikely to be associated with the ISM in the galaxies (Section \ref{sec:sed}).  Finally, we characterize the populations of the sources in these different systems in Section \ref{sec:prop}.




\section{Data}
\begin{figure*}
\centering
\plotone{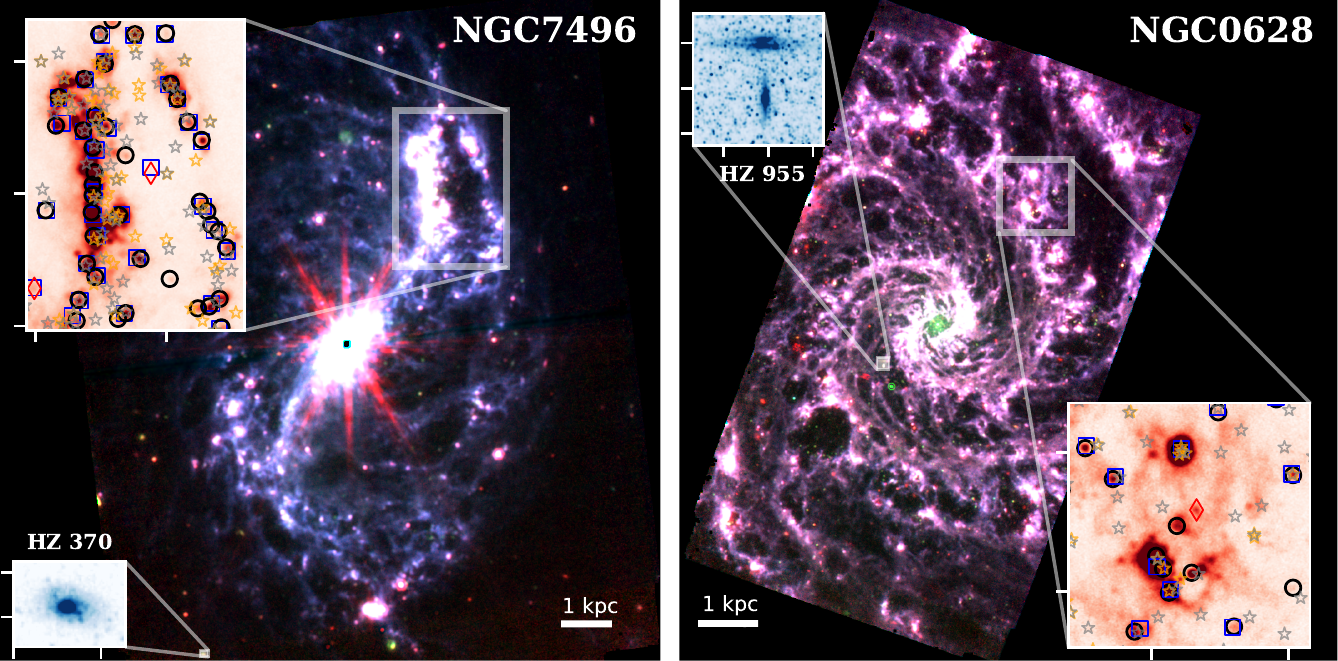}
\plotone{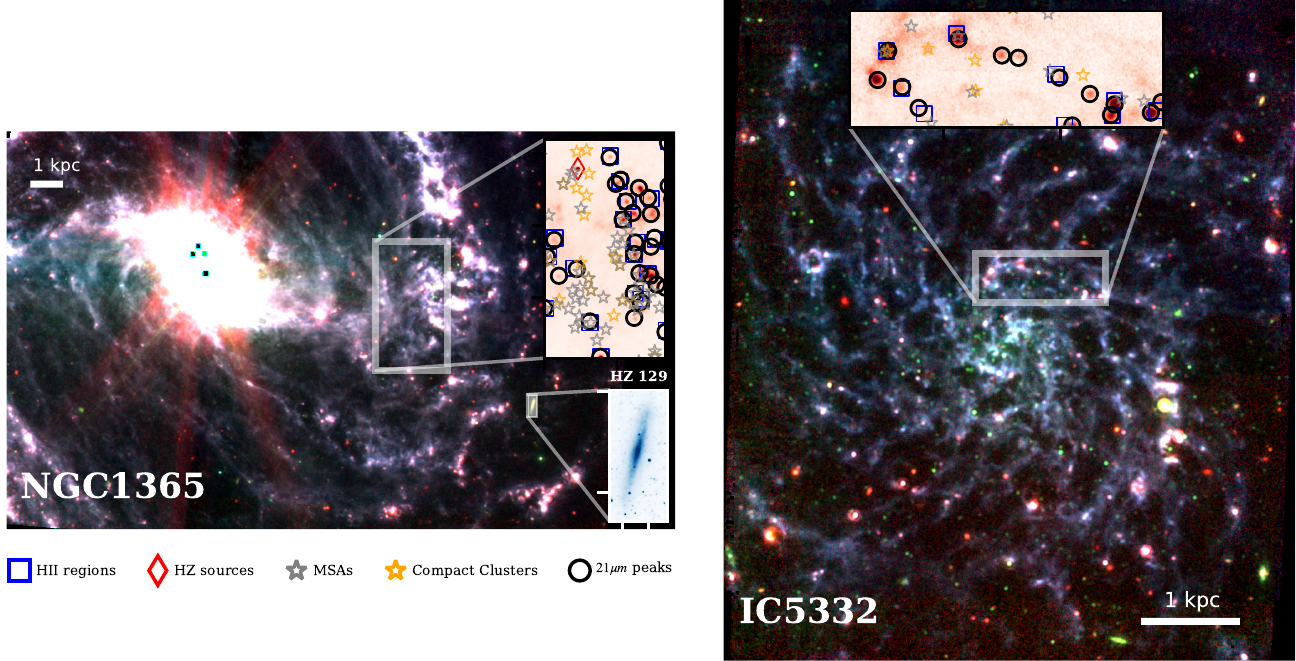}
\caption{RGB maps of NGC~7496 (top left), NGC~0628 (top right), NGC~1365 (bottom left), and IC~5332 (bottom right) at wavelengths of 21~$\um$ (red), 10~$\um$ (green) and 7.7+11.3~$\um$ (blue). The red zoom-in windows show peaks at 21~$\um$ (black circles), background galaxies (red diamonds), HII regions (blue squares), \edit1{multiscale} stellar associations (gray stars), and compact stellar clusters (orange stars).  In total 1271 sources are found throughout the images but we only show symbols in the inset images to reduce crowding.  We also highlight some representative background galaxies in the blue zoom-in windows, showing the F200W data at its native resolution.}
\label{fig:resscalingspitzer}
\end{figure*}

We analyze MIRI data from four galaxies in the PHANGS-JWST survey described by \citet{LEE_PHANGSJWST}: IC\,5332, NGC\,0628, NGC\,1365 and NGC\,7496 \edit1{with properties summarized in Table \ref{tab:props}}.  The data presented in this paper were obtained from the Mikulski Archive for Space Telescopes (MAST) at the Space Telescope Science Institute\footnote{The specific observations analyzed can be accessed via \dataset[10.17909/9bdf-jn24]{http://dx.doi.org/10.17909/9bdf-jn24}}.
We focus our analysis on the sources detected in the MIRI 21 \um data (F2100W filter, FWHM: $0\farcs67$) though we consider all data from the PHANGS-JWST filter set, which includes NIRCam (F200W, F300M, F335M, F360M) for NGC\,0628, NGC\,1365 and NGC\,7496, and MIRI (F770W, F1000W, F1130W, and F2100W) for all four targets\footnote{IC\,5332 NIRCam observations have not been obtained yet.}. \edit1{\citet{LEE_PHANGSJWST} describe the data reduction process including modifications of the default JWST pipeline and post-processing for the final images.  Of note, the modifications correct the processing for strong $1/f$ noise in the NIRCam data, correct for off-galaxy MIRI background images where available, and match background levels between individual MIRI fields while anchoring the overall background levels to low resolution archival data.}
This survey is being carried out in the context of the broader PHANGS survey, and we include maps of molecular gas content from the Atacama Large Millimeter/submillimeter Array (ALMA) $^{12}$CO(2-1) integrated intensity maps presented in \citet{phangs-alma}, the H$\alpha$ and H$\beta$ maps extracted from observations with ESO's Very Large Telescope (VLT) using the Multi Unit Spectroscopic Explorer \citep[MUSE;][]{phangs-muse}, and broadband optical imaging data from the {\it Hubble} Space Telescope (HST) as presented in \citet{phangs-hst}.  Figure \ref{fig:resscalingspitzer} shows the MIRI images of our four targets.

\begin{deluxetable*}{lcccccc}
\tablecaption{Summary of Galaxy Properties and Observational Parameters \label{tab:props}}
\tablehead{\colhead{Galaxy}  & \colhead{$D$}\tablenotemark{a} & \colhead{$\log_{10}(M_\star/M_\odot)$}\tablenotemark{b} & \colhead{$R_e$}\tablenotemark{b} & \colhead{$\ell_\mathrm{F2100W}$}\tablenotemark{c} & \colhead{$\ell_\mathrm{ALMA}$}\tablenotemark{c}& \colhead{$\ell_\mathrm{MUSE}$}\tablenotemark{c} \\
\colhead{} & \colhead{(Mpc)} & \colhead{} & \colhead{(kpc)} & \colhead{(pc)} & \colhead{(pc)} & \colhead{(pc)} }
\startdata
    IC 5332  & \phn9.01 & \phn9.7 & 3.6 & 29 & \phn32 & \phn38\\  
    NGC 0628 & \phn9.84 &    10.3 & 3.9 & 32 & \phn53 & \phn44\\
    NGC 1365\tablenotemark{d} &    19.57 &    10.0 & 2.8 & 64 &    130 & 110   \\
    NGC 7496\tablenotemark{d} &    18.72 &    11.0 & 3.8 & 61 &    150 & \phn77\\
\enddata
\tablenotetext{a}{Distances to targets based on observations aggregated in \citet{anand21} and \citet{shaya17}.}
\tablenotetext{b}{Galaxy properties including stellar mass ($M_\star$) and effective radius ($R_e$) based on the PHANGS compilation presented in \citet{phangs-alma}.}
\tablenotetext{c}{Projected linear resolutions at distances of the different galaxies based on observations with JWST ($0\farcs67$) and the varying resolutions of data from PHANGS-ALMA \citep[$0\farcs7$ to $1\farcs7$;][]{phangs-alma} and PHANGS-MUSE \citep[$0\farcs8$ to $1\farcs2$;][]{phangs-muse}.}
\tablenotetext{d}{NGC 1365 and NGC 7496 host IR-bright active galactic nuclei that saturate several of the JWST images.}
\end{deluxetable*}

\section{Compact Source Identification and Photometry} \label{sec:source}
To extract compact sources from the MIRI 21 \um data, we use the \textsc{astrodendro} software package \citep{astrodendro}.  We adopt this approach instead of a point source (star) identification approach because of the extended diffuse structure present throughout the maps (Figure \ref{fig:resscalingspitzer}) and because the ISM-tracing nature of the 21 \um band can lead to irregular morphologies even for compact sources. The highly structured shape of the F2100W PSF adds some additional complexity that confounds many point source identification algorithms. Dendrograms provide a non-parametric description of the contour structure in the emission \citep[][]{rosolowsky08}, which we then filter to identify our sources of interest.  

We first estimate the $1\sigma$ noise level in the map by inferring the difference in brightness between data at the 2.5th percentile in brightness compared to the 16th percentile in brightness, which would correspond to the interval between the $-2\sigma$ and $-1\sigma$ interval in a Gaussian distribution, leading to noise levels $\approx 0.3~\mbox{MJy~sr}^{-1}$.  This noise interval is likely an overestimate since even these percentiles of brightness distribution may contain real flux. We resort to these empirical measures since the pipeline noise maps in the First Results data include the effects of the MIRI coronagraph pixels, which are not actually included in the real maps.  We assume a single value of noise characterizes the whole map, though there is some spatial inhomogeneity in the noise structure at the 30\% level. 

We mask the data, and consider only regions above $5\sigma$ in brightness. Then, considering the region within the emission mask, we generate a dendrogram representation of the data, which finds local maxima in the map and identifies the contour levels below which each pair of local maxima is connected. We select local maxima that are $>2\sigma$ above the contour level at which that maximum merges with another maximum.  We further require that the maxima be spatially separated by at least one half width of the PSF at 21 \um ($0\farcs33$).  This approach yields a set of local maxima that are well defined and significant with respect to the \edit1{local} background.  This approach also identifies the diffraction structure from the PSF around bright sources as separate sources, so we further filter this set of local maxima by computing a roundness statistic around each peak inspired by {\sc daophot} \citep{daophot}.  Our statistic computes the absolute difference between a $2''$-square subimage and a version of the subimage that has been rotated by 180$^\circ$ around the local maximum.  We compare the sum of the absolute difference to the sum of the image, rejecting all local maxima for which this ratio is greater than 2, with this threshold value being chosen to eliminate diffraction features.

For each source, we measure the broad-band optical-IR spectral energy distribution (SED) using the JWST and HST data.  We convert the HST maps to surface brightness units (MJy~sr$^{-1}$) and use circularized convolution kernels generated using the methods in \citet{aniano11} to match the HST data to the F2100W resolution ($0\farcs67$). We sample each map at the location of the 21 \um local maximum, and we subtract an estimate of the median background calculated in an annulus with radii between $2\times$ and $3\times$ the width of the PSF. Note that these background levels are somewhat sensitive to the size of the annulus selected because of the strongly varying backgrounds and the possibility that the 21 \um peaks are extended.  \edit1{We estimate the local surface brightness uncertainty as the standard deviation of the data in this annulus, calculated using a median absolute deviation based estimator.  In the F2100W band, this leads to uncertainties from 4.1 to 16 $\upmu$Jy (16th to 84th percentiles).} We will revisit this approach in generating a more robust catalog in future work spanning the full PHANGS-JWST sample.

Contrasting the peak intensity with this local background estimate, we also reject any source where the surface brightness at F2100W has a peak-to-background ratio $<5$. The net effect of these filters is to find point-like and slightly resolved bright sources in our images with typical diameters $\lesssim 2\times 0\farcs67$.  

We convert surface brightness to flux density by measuring the solid angle subtended by the MIRI PSF in the F2100W band ($\Omega_\mathrm{F2100W} = 1.61\times 10^{-11}~\mathrm{sr}$), calculated using the models from the {\sc WebbPSF} package \citep{webbpsf}.  We note that we complete our analysis at a common angular resolution but distances to our targets vary by a factor of two.  Hence, our analysis implicitly includes changing physical scales \edit1{(Table \ref{tab:props})} and different minimum recoverable luminosity between the different objects, and we note where this effect shapes our results below.

The AGN at the centers of NGC\,1365 and NGC\,7496 saturate the detector and scatter light over a wide area in the map.  To identify structures that could be caused by this scattered light, we match the F2100W PSF models to the location of the AGN and the orientation of the diffraction pattern, rejecting sources where the PSF amplitude is $>0.1\%$ of the PSF maximum.  This ends up covering a region of $\sim 15''$ in diameter ($\sim 1.3$ kpc in these targets).  Some other bright sources in the center of NGC\,1365 are also saturated and not included in this analysis though these are further explored in \citet{SCHINNERER_PHANGSJWST} and \citet{LIU_PHANGSJWST}. 

Our final catalog consists of 1271 compact sources with 188 in IC\,5332, 502 in NGC\,0628, 367 in NGC\,1365 and 214 in NGC\,7496. In addition to the JWST and HST-based SEDs, we also sample the MUSE H$\alpha$ and H$\beta$ maps and the ALMA CO(2-1) integrated intensity and line width maps at the location of the 21 \um local maxima. For the ALMA and MUSE measurements, we do not perform any background subtraction. These maps are at coarser resolution (up to $1\farcs 7$ depending on band and target; \edit1{see Table \ref{tab:props}}) and thus yield slightly smoothed estimates of the H$\alpha$, H$\beta$, or CO(2-1) that would be measured at the JWST resolution.  \edit1{The subsequent  CO-based analysis does not strongly depend on the differences in resolution, but measurements of H$\alpha$ luminosity may be overestimated and could affect our results.  In the case of a flat H$\alpha$ map over the MUSE PSF area, the H$\alpha$ brightness could be overestimated relative to the JWST beam by the ratio of PSF areas, which could be up to a factor of 3 in this extreme case (NGC 1365). However, the galaxies show similar resolutions between MUSE and JWST so even for the extreme case, the overestimate will be less severe.}


\begin{figure*}
\centering
\includegraphics[width=\textwidth]{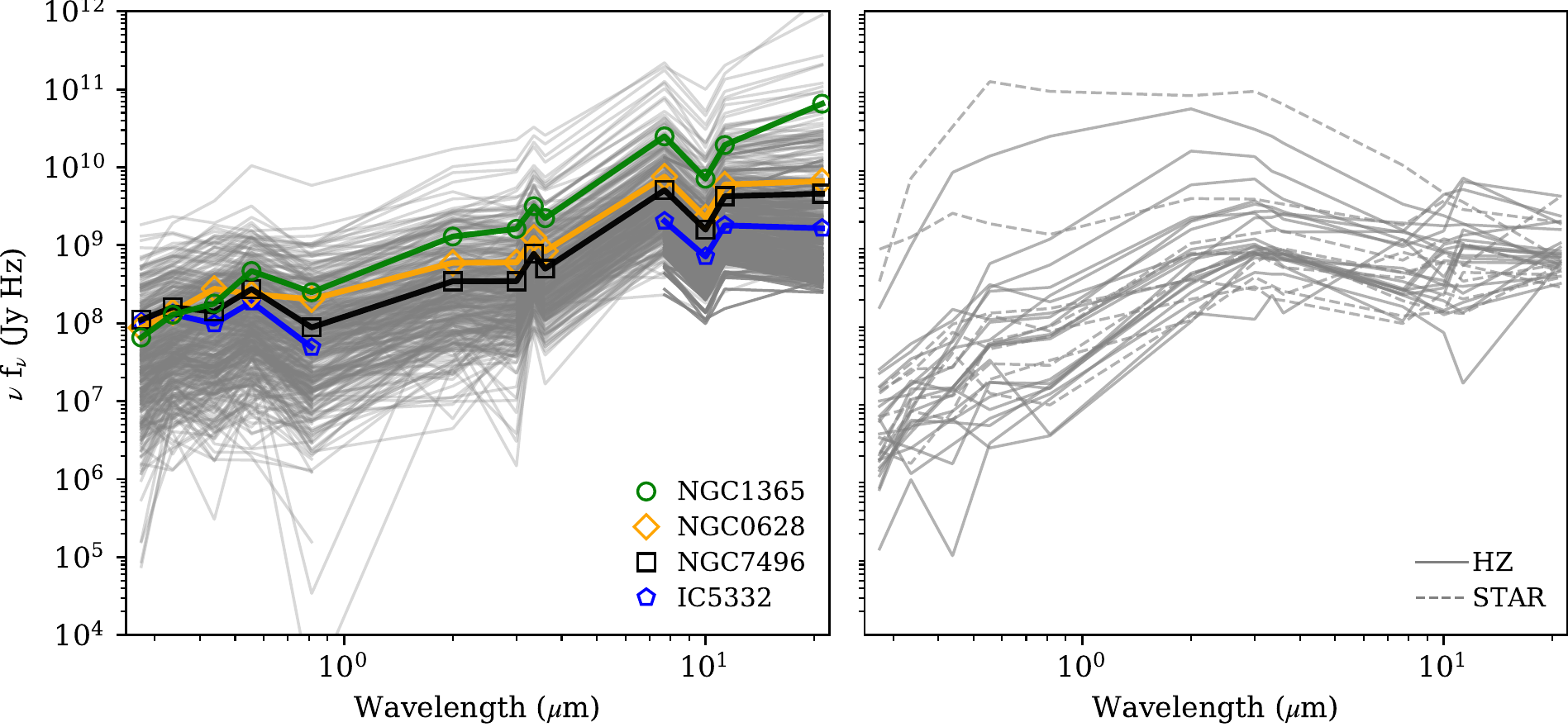}
\caption{The SED of ISM sources (left) and eye-confirmed background galaxies \edit1{and} stars (HZ \edit1{and} STAR; right). We only consider the sources that have $>5\sigma$ detections in all (JWST+HST) bands: 384 ISM sources and \edit1{30} HZ or STAR sources.  The grey lines represent the SED of each object, and the solid colored lines show the mean of all SEDs for each individual galaxy.}
\label{fig:seds}
\end{figure*}

\section{Spectral Energy Distributions} \label{sec:sed}

\begin{figure*}
    \includegraphics[width=\textwidth]{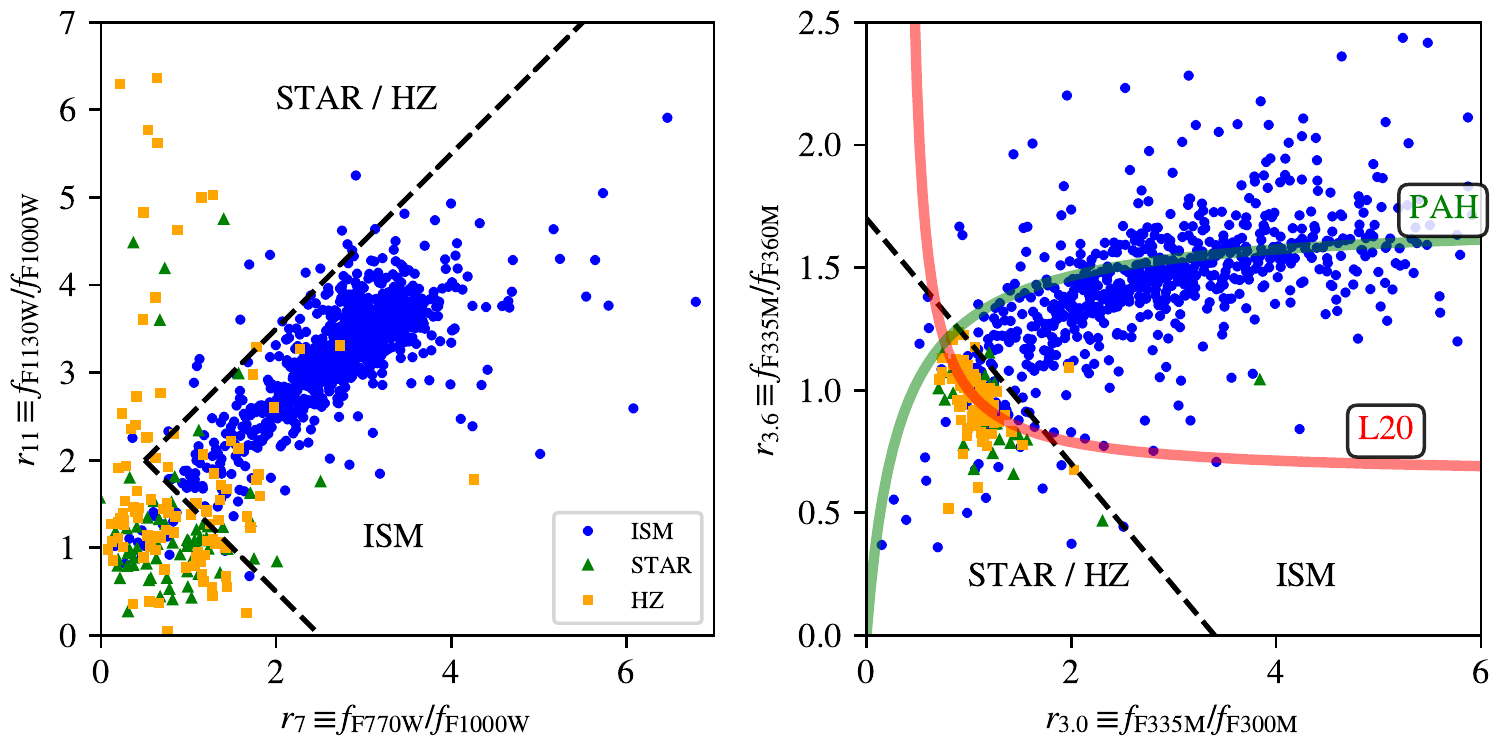}
    \caption{Color-color plots of the 21 \um selected sources in the MIRI (left) and NIRCam (right) bands constructed from band ratios that highlight significant PAH emission for $z=0$ sources.  Dashed lines indicate loci that partition the color-color space into sources with bright ISM emission \edit1{(ISM)} vs.\ background galaxies \edit1{(HZ)} and extreme AGB stars that lack a PAH feature \edit1{(STAR)}.  Classifications are made by eye in examining the full SED and morphology of sources in the full PHANGS data set. The L20 line indicates the proposed line for stellar colors from \citet{lai20}.  The green line in the right-hand plot indicates the expected flux ratios for PAH emission in \citet{SANDSTROM2_PHANGSJWST}.}
    \label{fig:colors}
\end{figure*}

Of the 1271 compact sources, 1083 have NIRCam as well as MIRI data (recall that IC\,5332 currently lacks NIRCam imaging).  Figure \ref{fig:seds} illustrates the SEDs for our sources.  Most sources are bright at 21 \um by selection, with a characteristic ``dip'' at 10 $\um$ defined by the gap between the bright PAH features at 7.7 \um and 11.3 \um.  Sources with NIRCam imaging also show the PAH feature at 3.3 \um as a ``bump'' in the SED. \edit1{There is modest variation in the median SEDs between sources in different galaxies, and the bright sources in the nuclear star forming ring of NGC\,1365 increase the median flux profile for that galaxy.} 

To better understand the nature of the identified sources, two authors (HH and ER) visually examined the SEDs and an atlas of multiwavelength images showing the full set of native resolution JWST, HST, ALMA and MUSE data for each source.  We find that most sources appear to be star forming regions associated with slightly resolved clusters or associations of reddened stars, bright PAH features, and CO or H$\alpha$ emission. We refer to these sources as ``ISM'' sources because their SED reflects the presence of strong PAH features, which emerge from dust mixed with ISM material.
These ISM sources are typically embedded in diffuse filamentary structure visible in the F2100W, F1130W, and F770W bands \citep{THILKER_PHANGSJWST}. 

In Figure \ref{fig:seds}, we also show the SEDs of the visually-identified non-ISM sources.  Several of these appear to be dusty stars, likely extreme-AGB stars \citep[][Thilker et al., in prep.]{corbelli11}, which are recognizable by a smooth SED in the near- and mid-IR with weak PAH features and a point-like morphology in F200W. We designate these as ``STAR'' sources.  The remaining sources appear to be background galaxies. These have an extended shape in F200W images and no PAH features, likely because the PAH features are redshifted out of the corresponding JWST bands.  We refer to these as high-$z$ or HafeZ\footnote{in inspiration from the Persian poet Hafez.} sources (hereafter HZ). 


Leveraging these visual classifications, we attempt to identify color cuts that could effectively achieve the same assignments using ratios among the MIRI and NIRCam bands. Given the prominence of PAH features in the ISM sources and their absence or weakness in the HZ and STAR sources, we focus on the brightness of the bands with strong PAH features relative to the nearby PAH-free bands. The resulting color comparisons are illustrated in Figure~\ref{fig:colors}. In the left panel, we plot $r_7\equiv f_\mathrm{F770W}/f_\mathrm{F1000W}$ vs. $r_{11} \equiv f_\mathrm{F1130W}/f_\mathrm{F1000W}$, which normalizes the bright PAH features in the F770W and F1130W bands by the nearby F1000W band. F1000W is expected to reflect mostly continuum or silicate absorption.  ISM-like features with strong PAH emission fall along a tight locus with $r_{11} \approx 1.25r_7$.  
These ISM features clearly separate from the STAR/HZ sources in this space and we define a boundary
\begin{eqnarray}
r_{11} & > & 2.5 - r_7, \\
r_{11} & < & r_7 + 1.5,
\end{eqnarray}
such that nearly all sources in this region are classified as lines of sight dominated by the ISM emission.  Sources outside this region lack a dip in the 10 \um relative to the 7.7 \um and 11.3 \um PAH features that is characteristic of typical ISM emission from local star forming galaxies. 

We construct a similar color-color diagram using the NIRCam bands (Figure \ref{fig:colors}, right). Here we focus on identifying lines of sight with bright PAH emission by contrasting the F335M band, which covers the 3.35 \um PAH feature, against the more continuum-dominated nearby F330M and F360M.  For $r_{3.0} \equiv  f_\mathrm{F335M}/f_\mathrm{F300M}$ and $r_{3.6} \equiv  f_\mathrm{F335M}/f_\mathrm{F360M}$, the regions with a significant PAH feature show 
\begin{eqnarray}
r_{3.6} &>& 1.7 - 0.5 r_{3.0},
\end{eqnarray}
which appears as a dashed line in the right panel of Figure \ref{fig:colors}. For reference, we also show the line defined by \citet{lai20} that indicates the locus of stellar continuum in the JWST bands and the locus of PAH emission defined by \citet{SANDSTROM2_PHANGSJWST} \citep[see also][]{RODRIGUEZ_PHANGSJWST}.    

Our partition of this color space generally yields an accurate match to our manual source classification.  The MIRI-based diagnostic (Figure \ref{fig:colors}, left) is 94\% accurate on the 1271 sources with measured MIRI colors. Here accuracy is defined as the fraction of correct classifications (true negatives and true positives) compared to the by-eye SED and image classification.  There still remain notable false positives and negatives, which stem partly from complex backgrounds affecting the extracted flux values. The NIRCam-based partition is only 88\% accurate on the 1083 sources with NIRCam data (recall again that IC\,5332 lacks \edit1{NIRCam} data).  This lower accuracy results in substantially more false negative classifications using NIRCam colors (128) than using MIRI colors (34). These false negatives are sources identified as not being ISM features because of the lack of a strong 3.35 \um PAH feature but that appear to be ISM sources in our manual assessment. If we classify a source as being an ISM feature based on having at least one diagnostic consistent with the color cuts, the accuracy of the classification rises marginally to 95\% for the 1083 sources with both MIRI and NIRCam measurements. \edit1{Misclassification could also occur when there is significant PAH (or ISM) emission in the bands we are using as ``continuum'' bands: F1000W, F300W and F360W. For example, \citet{lai20} note the 3.47 \um PAH band may contribute to the F360W light, which is also seen in the PHANGS-JWST sample \citep{SANDSTROM2_PHANGSJWST}.  \citet{LEROY1_PHANGSJWST} show evidence that F1000W may have a significant contribution from the wings of the PAH features in adjacent bands. The 9.7 \um silicate feature is often found in absorption in evolved stars \citep[e.g.,][]{werner80} which could make these stars appear more ISM-like.}

To maximize the accuracy of our assessment on the largest number of sources while retaining the ability to apply these results to other targets, we proceed using the MIRI-based diagnostic as the primary source classifier.  Future work will rely on such color classification since the number of sources will be too large for human classification with continued JWST operations.

While our compact source catalog contains \edit1{187} non-ISM sources from background galaxies and stars, these sources do not contribute significantly to the 21 \um flux density of our targets.  Using the MIRI classifier, we find that 97\% of the total F2100W flux density in our catalogs comes sources classified as ``ISM'' across all targets.  

\section{Properties of the 21 \um Sources}
\label{sec:prop}

Young massive clusters and associations are thought form in dark clumps of gas that the newly formed stars then disrupt with winds and radiation.  There should thus be a phase of star formation where the young stars are embedded and visible primarily in the IR before they break out and become visible in the optical. This picture has been borne out in other systems \citep[e.g.,][]{kim21}. Our data broadly support this picture: from the visual inspection, the compact 21-\um bright sources detected by JWST in these targets appear to be primarily star-forming regions hosting high-mass star formation, though relatively few sources are associated with a truly optically-dark phase \citep[see also][]{WHITMORE_PHANGSJWST, KIM_PHANGSJWST}.  In this section, we quantify this impression by examining the associations between the 21 \um sources and other star formation tracers and then examining embedded sources. For this analysis, we restrict our analysis to 1085 compact sources that are classified as ``ISM'' by the MIRI-based flux ratio criterion.


\subsection{SED Fitting and Derived Properties}
\label{sec:sedfit}
\begin{figure}
    \centering
    \includegraphics[width=\columnwidth]{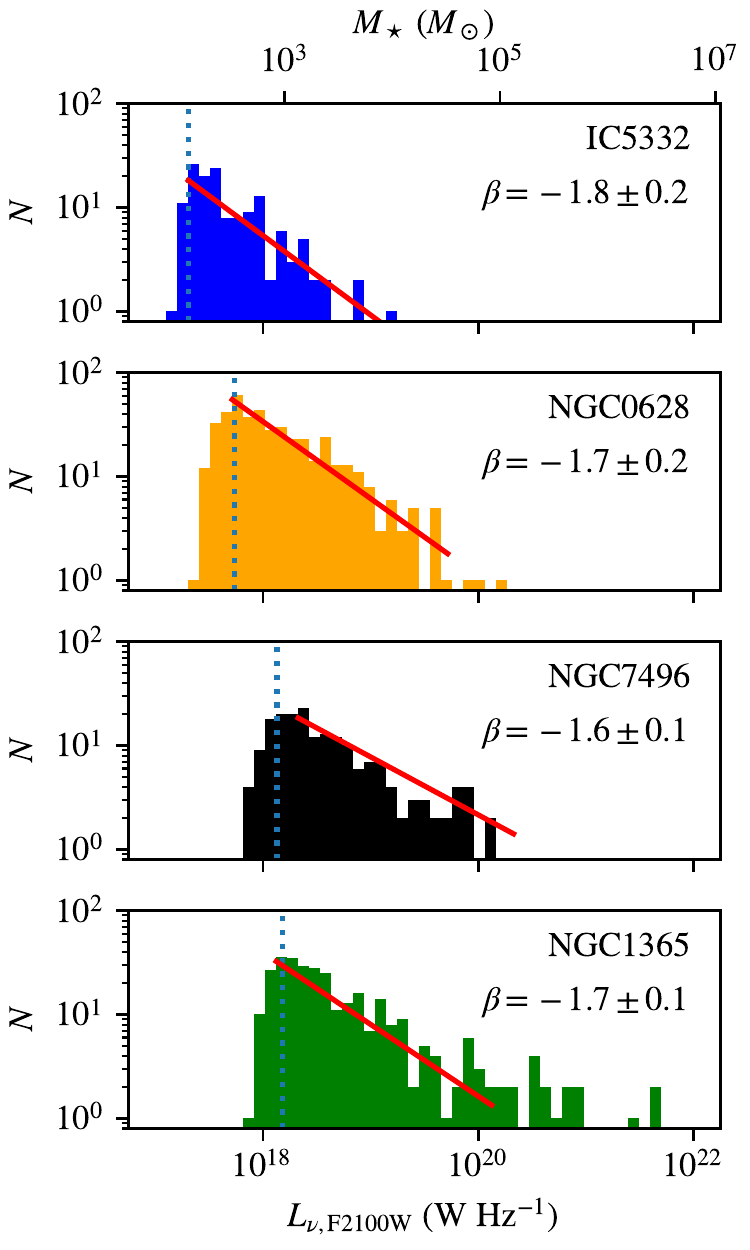}
    \caption{Luminosity distribution of 21 \um sources classified as ISM sources in the JWST MIRI data.  The distributions are grouped by host galaxy in order of increasing distance (top to bottom). The blue dotted line indicates the mean $5\sigma$ error in source luminosity. The red lines indicate a fit to the power-law distribution with $dN/dL_\nu \propto L_\nu^{\beta}$ over two orders of magnitude. The value of $\beta$ is labeled in each panel.  The mass scale on the top axis translates the luminosity measurements to equivalent masses using the median ratio we determine from SED fitting (Section \ref{sec:sedfit}).}
    \label{fig:lumdist}
\end{figure}

To characterize these compact sources, we use the \textsc{cigale} code \citep{cigale} to estimate the age and mass of stellar populations that could yield our measured JWST and HST luminosity densities, as well as some properties of the surrounding photodissociation region. For the fit, we consider a double exponential star formation history with an $e$-folding time of ranging from 0.01 to 10\,Myr, and a late starburst with the $e$-folding time ranging from 0.001 to 0.1\,Myr.  Based on the characteristic timescales for star formation in our galaxies \citep[e.g.,][]{chevance22}, we consider a range of stellar population ages from 3 to 25\,Myr.  Based on the the Balmer decrement analysis of H$\alpha$ and H$\beta$ emission \citep[][Belfiore et al., subm.]{phangs-muse, santoro22, GROVES_HIICAT}, we restrict our models to $0.1<E(B-V)<3$, and we use a \citet{calzetti00} attenuation law. We include a \citet{draine14} dust model with a mass fraction of PAHs ranging from 0.47\% to 7.32\%. Since individual \ion{H}{2} regions can have intense radiation fields, we consider $0.5 < U_\mathrm{min}/U_0< 50$, where $U_\mathrm{min}$ is the parameter from \citet{dl07} that characterizes the interstellar radiation field (ISRF) and $U_0$ is the ISRF in the solar neighborhood \citep{mmp83}. We consider a fraction of the dust mass heated by the power-law distribution of radiation fields with intensity up to $U_\mathrm{max}=10^7\,U_0$ with a fraction of $\gamma=0.1$. Furthermore, we allow the exponent of power-law distribution to span a range of 1 to 3 in intervals of 0.5. Finally, we choose a Chabrier Initial Mass Function (IMF) for our model.  We summarize our results in Figures \ref{fig:lumdist} and \ref{fig:hacorr}. Our fits typically find that the 21 \um sources have SEDs consistent with stellar population masses $10^2 < M_\star/M_\odot < 10^{4.5}$.  We find a range of mass-weighted ages spanning 2 to 25 Myr though most objects are $<8$~Myr.   Based on the \textsc{cigale} fits, we find a median ratio of $\nu L_\nu / M_\star = 98~L_\odot M_\odot^{-1}$ for our sources with a variation of $\pm 0.3$~dex.

Figure \ref{fig:lumdist} shows the luminosity distribution of sources in the four galaxies, ordered by distance to the target.
All distributions show a turnover at the low-mass end that is due to completeness since the value corresponds well to the typical $5\sigma$ luminosity uncertainties  (vertical dotted lines in Figure \ref{fig:lumdist}). The noise uncertainty is estimated from the typical errors in the flux measurements and thus includes both image noise and the complexity of the background. Above this limit, the distributions show good agreement with a power-law distribution function (red line) shown in the Figure with a form of $dN/dL_\nu \propto L_\nu^{\beta}$ running over two orders of magnitude in source luminosity. We use the \edit1{median} $\nu L_\nu / M_\star$ ratio from the \textsc{cigale} fits to define the top axis of the graph.   We fit each binned distribution with a simple linear regression ranging from the $5\sigma$ limit to 100$\times$ this value to provide an estimate of the index of the distribution.  We assess uncertainties by rebinning the data to different widths and offsets and refitting the distribution. In NGC 1365, there is a tail of sources at high luminosity, which consists of the sources in the nuclear star-forming ring of this system that are not included in the fit.  These high mass, young sources are explored further in \citet{WHITMORE_PHANGSJWST}.




Our derived values of the power-law index (typically $\beta \approx -1.7 \pm 0.1$) are slightly shallower (top-heavy) compared to standard analyses of star cluster masses, which find $\beta=-2.0\pm 0.3$ \citep{mok20}, though analyses of young clusters sometimes find shallower indices \citep{whitmore99, adamo17, adamo20} for the cluster mass function.  The slopes are essentially the same as \ion{H}{2} region luminosity function in PHANGS targets \citep[$\beta_{\mathrm{HII}}=-1.7$,][]{santoro22}. Though careful source identification and detailed SED fitting remain the most precise way to assess the cluster mass function, the overall agreement between different results and reasonably stable mass-to-light ratio suggest that similar to \ion{H}{2} region luminosities, the IR peaks may provide a useful indicator of the mass distribution of young associations and clusters.  If we adopt a constant scaling of $\nu L_\nu / M_\star$, the slope of the luminosity distribution would also trace the mass distribution of these objects. These values can be compared to the mass distribution of clusters from optical analyses though we are not directly selecting for long-lived bound objects. 



\subsection{Association with Star Formation Tracers}
\label{sec:association} 
The 21 \um compact sources are tightly correlated with \ion{H}{2} regions. Over $92\%$ of sources in the portion of the JWST images that overlaps with the MUSE field of view are associated with an \ion{H}{2} region from the catalog by \citet{santoro22} and \citet{GROVES_HIICAT}, which was generated using the \textsc{HiiPHOT} software \citep{hiiphot}.  We define association as an \ion{H}{2} region having any of its pixels within $0\farcs8$ of the catalog position of a 21 \um source. This association rate far exceeds the fraction that would be expected for random placement of the sources within the field of view ($11\%$). As \textsc{HiiPHOT} finds compact \textsc{Hii} regions (ignoring diffuse H$\alpha$ emission), this high rate of coincidence shows that nearly all 21 \um bright sources are also associated with an optically visible tracer of high-mass star formation.  The reverse is not true: only 15\% of \ion{H}{2} regions have associated 21 \um sources. This lower fraction likely stems from the 21 \um source catalog having fewer objects and these objects being preferentially associated with the brightest H$\alpha$ sources (see also \edit1{Section} \ref{sec:hacorr} below).  We find that the median H$\alpha$ flux of \ion{H}{2} regions associated with 21 \um sources is $\approx 4\times$ the median flux of the \ion{H}{2} region catalog.

Similarly, these 21 \um sources are also associated with a young, high-mass stellar population seen in the optical.  Specifically, 74\% of 21 \um sources in the region observed by both HST and JWST are found within $0\farcs8$ of a source in the multiscale stellar association (MSA) catalog of \citet{LARSON_MSA}. The MSA catalog finds associations of nearby bright compact sources on a specific spatial scale \edit1{where we use 64 pc scale associations, which are chosen for similarity to the JWST resolution in our most distant targets.  The MSA analysis} then uses \textsc{cigale} fits to \edit1{HST} multiband photometry to determine stellar ages and masses.  While lower than for \ion{H}{2} regions, this rate still indicates close association with the optically-visible stellar population since a random association would only produce a rate of 26\%. The 15\% of stellar associations that are associated with 21 \um sources have a geometric mean age of 7 Myr, which is younger than the population of regions as a whole (12 Myr). 

We also compare to the \edit1{PHANGS-HST} cluster catalog \citep{brad, phangs-hst, thilker22, deger22}, which finds compact stellar clusters in the HST data and measures the the ages and masses of those objects.  We consider the objects associated if they are offset by $<0\farcs 8$ on the plane of the sky and find stellar clusters associated with 15\% of the 21 \um sources.  This result indicates a weaker correlation since random placements only lead to 9\% of sources being associated. The geometric mean of associated cluster ages is 4 Myr, significantly younger than the geometric mean age of the full catalog (30 Myr). Thus, the 21 \um sources are also associated with a young, optically visible stellar component though the association is not as strong as with the \ion{H}{2} regions.   The weaker correlation likely arises because bound clusters result from only a fraction of star formation and also some selection effects against young clusters from, e.g., dust extinction.  Studying embedded clusters with JWST \citep[e.g.,][]{RODRIGUEZ_PHANGSJWST, WHITMORE_PHANGSJWST} offers a good opportunity for resolving these possibilities.


\subsection{A Dearth of Embedded Sources}
\label{sec:hacorr}

\begin{figure*}
    \centering
    \includegraphics[width=\textwidth]{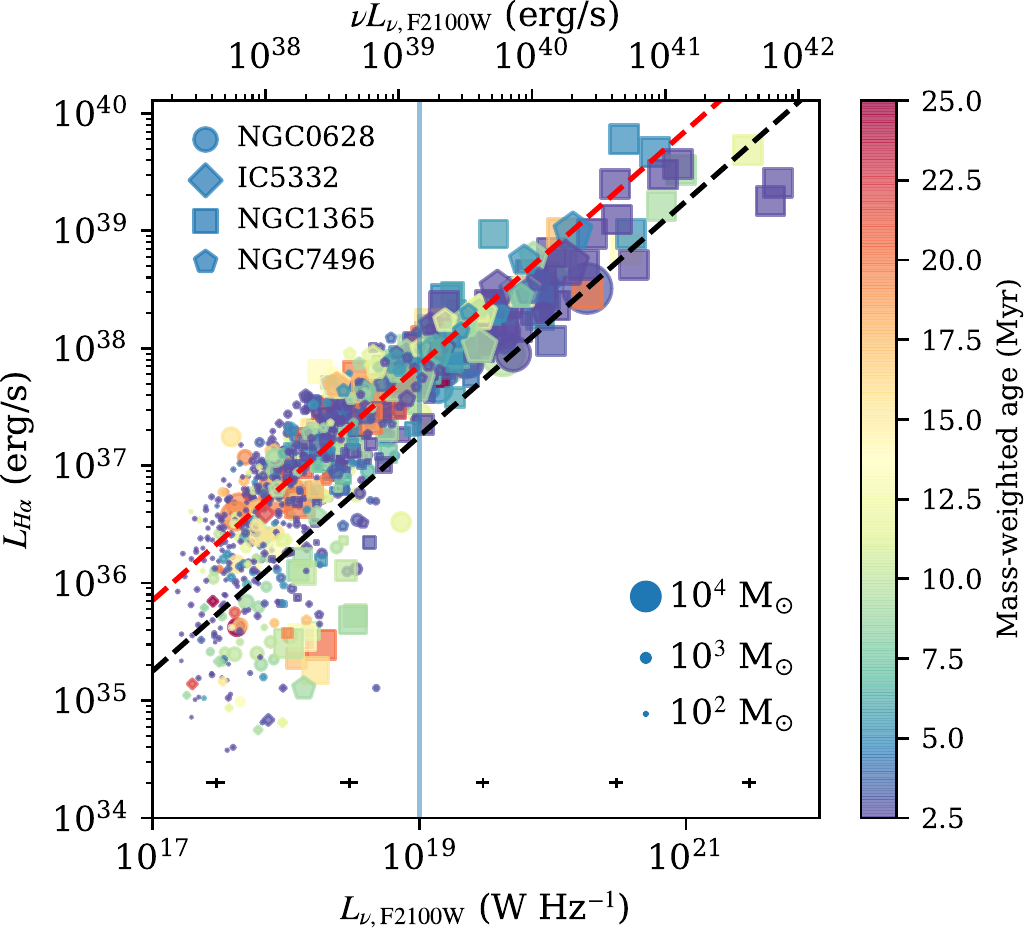}
    \caption{Correlation between luminosity in the extinction corrected H$\alpha$ line as a function of 21 \um luminosity density. The color bar shows age of the sources and the total stellar mass of each region is represented by its size. Both age and mass are estimated via SED fitting using \textsc{cigale} (Section \ref{sec:sedfit}). Most regions show an excellent correlation between 21 \um and H$\alpha$ luminosity.  The black dashed line indicates the threshold below which we regard a 21 \um source as being H$\alpha$-faint \edit1{($L_\mathrm{H\alpha} / \nu L_{\nu, \mathrm{F2100W}}=0.05<1/80$)}, and the red line indicates the mean $L_\mathrm{H\alpha} / \nu L_{\nu, \mathrm{F2100W}}=0.05$.  The vertical line at $L_\nu = 10^{19}~\mathrm{W~Hz^{-1}}$ is the threshold below which we search for embedded sources. \edit1{We plot representative fractional error bars in the bottom of the plot in five different $L_{\nu, \mathrm{F2100W}}$ ranges.}}
    \label{fig:hacorr}
\end{figure*}

There is an excellent correlation between the 21 \um luminosity of a source and the luminosity of the H$\alpha$ line at the source location \citep[see also][]{LEROY1_PHANGSJWST}. In Figure \ref{fig:hacorr}, we plot the correlation between the luminosity of the sources in the F2100W and the extinction-corrected H$\alpha$ luminosity calculated using the Balmer decrement method applied to the MUSE H$\alpha$ and H$\beta$ maps (Belfiore et al., subm.).  The H$\alpha$ luminosity is calculated by sampling the H$\alpha$ flux maps and scaling by the projected area of the JWST PSF: $L_\mathrm{H\alpha} = f_\mathrm{H\alpha} \Omega_\mathrm{F2100W} d^2$. The figure shows a nearly linear correlation between these two tracers throughout our sample with $\langle L_\mathrm{H\alpha} / \nu L_{\nu, \mathrm{F2100W}}\rangle = 0.05$.  A linear relation describes the population above $L_{\nu, \mathrm{F2100W}}\approx 10^{19}~\mathrm{W\ Hz^{-1}}$; but below this threshold, there is a population of sources that have significant 21 \um emission but minimal or no associated H$\alpha$ emission.  \edit1{We also investigated these results on a per-galaxy basis and found no significant differences in these behaviors despite the variations in luminosities sampled (Figure \ref{fig:lumdist}).} These sources are our candidates for embedded sites of star formation based on the 21 \um data.

We construct a boundary at  $L_\mathrm{H\alpha} / (\nu L_{\nu, \mathrm{F2100W}}) = 1/80$ (dashed black line in Figure \ref{fig:hacorr}) and examine the sources with $L_{\nu,\mathrm{F2100W}}<10^{19}~\mathrm{W~Hz^{-1}}$.  We find 105 of the 1085 (10\%) ISM-classified sources are ``embedded'' under these criteria. These sources are typically found in extinction features in the HST data \citep{THILKER_PHANGSJWST}. We find that the median aperture-subtracted flux densities in the HST bands are all typically negative at the $1\sigma$ to $2\sigma$ level, whereas sources with H$\alpha$ have positive optical fluxes on average.  Visual examination shows several sources overlap with bright CO clouds in the ALMA data ($\sim 1''$ resolution), however a comparable number of sources are found in small voids in the CO emission. This diversity of CO morphologies does not indicate a clear link to a specific phase of molecular cloud evolution.  We also examined whether these embedded sources are distinct in several JWST color spaces but found no significant differences between embedded and optically visible sources.
This result is similar to the results of \citet{kim21} and \citet{KIM_PHANGSJWST}, \edit2{who modeled the lifecycle of a typical star forming region in NGC 0628 and other nearby galaxies based on the location and scale dependence of the CO-to-21 \um and H$\alpha$-to-21 \um ratios. They found that the period of time when a region emits bright 21 \um emission only slightly precedes the time when it emits bright H$\alpha$ (specifically they inferred regions to be H$\alpha$-bright for 70\% of the total time they emitted bright 21 \um), which agrees well with our observations. Note, however, that the current analysis} does not apply a background subtraction to the extinction corrected H$\alpha$ maps, in contrast with the analysis of the \ion{H}{2} regions in Section \ref{sec:association} and the timescale analysis in \citet{KIM_PHANGSJWST}.  Hence, the presence of diffuse emission may affect \edit2{our} results.

%

Based on our \textsc{cigale} fits, we see only a few high-mass clusters ($>10^4~M_\odot$) in NGC 1365 that are deeply embedded (below the black dashed line).  These objects are explored more in \citet{WHITMORE_PHANGSJWST, SCHINNERER_PHANGSJWST, LIU_PHANGSJWST}.  For low mid-IR luminosities, all the young ($<10$~Myr), embedded sources have SED fits that also find low mass measurements ($M<10^{3}~M_\odot$), reaching levels where stochastic sampling of the IMF can become important \citep[][]{dasilva12} and a cluster may not form stars with significant ionizing flux. However, in smoothing the data to the relatively coarse resolution of the F2100W filter, we may obscure truly embedded sources by combining the light with brighter surrounding sources. Examining these sources in shorter wavelength and thus higher resolution data is another promising route for identifying embedded sources \citep{RODRIGUEZ_PHANGSJWST}.

In summary, we find that only a few ($<10\%$) sources visible at 21 \um that are candidate embedded star forming regions traceable primarily in the mid-infrared.  These sources are fainter than the overall population of 21 \um emitters, associated with young, low mass stellar structures ($M<10^3~M_\odot$, age $<$ 10~Myr) as a whole and contribute $<1\%$ of the 21 \um luminosity.  

\subsection{Spatial Associations with Gas vs. \ion{H}{2} regions}

As a final assessment of how the 21 \um sources are related to the different stages of the star formation process, we compare the spatial distribution of these sources to those of giant molecular clouds and \ion{H}{2} regions. We measure the median minimum distance between the 21 \um source and the centers of \ion{H}{2} regions in the MUSE-derived catalog \edit1{\citep{santoro22,GROVES_HIICAT}.}  We compare this to the median minimum distance between a given 21 \um source and the brightest point in a giant molecular cloud (GMC), using the PHANGS GMC catalogs \citep{rosolowsky21, hughes_catalog}.  This analysis uses only the overlapping regions in the JWST, ALMA and MUSE maps.  We use GMC catalogs extracted from the best resolution maps available in the ALMA data.  We compare these median minimum distance measurements to randomized catalogs of \ion{H}{2} regions and GMCs, where we create catalogs with the same average radial number density, number of sources, and spatial footprint as the true catalogs, but randomly distributed across the field of view.  We then measure the median minimum distance between 21 \um sources and center positions in the random catalog. 

\begin{deluxetable}{lcccc}
\tablecaption{Median values of the minimum offset between each of the 21 \um sources and the nearest \ion{H}{2} region ($d_\mathrm{IR-HII}$) or GMCs ($d_\mathrm{IR-GMC}$).\label{tab:offsets}}
\tablehead{\colhead{Galaxy}  & \colhead{$d_\mathrm{IR-GMC}$} & \colhead{$d_\mathrm{IR-HII}$} & \colhead{$d_\mathrm{IR-GMC}$}& \colhead{$d_\mathrm{IR-HII}$} \\
\colhead{} & \colhead{(pc)} & \colhead{(pc)} & \colhead{(pc)} & \colhead{(pc)} }
\startdata
& \multicolumn{2}{c}{Observed} & \multicolumn{2}{c}{Random\tablenotemark{a}} \\
\cmidrule(lr){2-3} \cmidrule(lr){4-5}
    IC 5332 & \nodata & 23 & \nodata & 109 \\
    NGC 0628 & 80 & 19 & 590 & 283 \\
    NGC 7496 & 140 & 27 & 590 & 280\\
    NGC 1365 & 220 & 60 & 830 & 880
\enddata
\tablenotetext{a}{Values for random catalogs of \ion{H}{2} regions and GMCs.}
\end{deluxetable}

Table \ref{tab:offsets} shows the median minimum offsets. In nearly all cases, we find the same basic pattern: 21 \um sources are clustered with both \ion{H}{2} regions and GMCs with respect to the random catalogs.  However, the separation between 21 \um sources and the nearest \ion{H}{2} region is systematically smaller than the separation between the source and the nearest GMC.  The median minimum offset for a 21 \um source and the nearest \ion{H}{2} region is typically smaller than the JWST or the MUSE point spread function projected to these galaxies.  This correlation aligns well with the analysis of separation between sources seen in \citet{KIM_PHANGSJWST}, which in that framework implies that the 21 \um sources and \ion{H}{2} regions overlap for a longer part of their lifetime than GMCs and IR sources. This also complements the results of \citet{LEROY1_PHANGSJWST} who find that the bright mid-IR map is well represented by the H$\alpha$ emission. In this simple analysis presented here, these results require some caution: the smaller offsets for \ion{H}{2} regions can be influenced by there being more \ion{H}{2} regions than GMCs.  While the ALMA maps are the best available tracers of the star forming molecular gas, the other two maps show far broader dynamic range in their brightness levels.  We do not report results for IC\,5332 GMCs since the ALMA map only recovers 11 GMCs in this low metallicity system.

\section{Conclusions and Summary}

We have identified and investigated the compact 21 \um source population in four nearby galaxies observed with JWST as part of the PHANGS project.  Using a dendrogram-based source identification algorithm, we identified 1271 sources in the survey area and measured their flux densities in the JWST bands (Section \ref{sec:source}). By using the characteristic signature of PAH emission to make color cuts on the MIRI and NIRCam photometric data, we are able to identify and exclude a contaminating population of dusty stars and background high-$z$ galaxies. This results in 1085 sources that show emission features consistent with being ISM in the target galaxies. For these ISM-like sources, we find:
\begin{enumerate}
    \item Multi-band SED fitting of these compact objects shows that they are associated with stellar structures (clusters, associations) with stellar population mass $10^2 < M_\star/M_\odot < 10^4$.  We find $\nu L_{\nu,\mathrm{F2100W}}/M_\star \approx 98~L_\odot / M_\odot$ with a range of $\pm 0.3$~dex.
    \item In each galaxy, the luminosity distribution is well described by a power-law relationship ranging over two orders of magnitude with form $dN/dL_\nu \propto L^\beta$ The power law ranges from $\beta=-1.8$ to $-1.6$ (Figure \ref{fig:lumdist}), which agrees well with the luminosity distribution of \ion{H}{2} regions and the mass distribution of young clusters in previous work.
    \item The 21 \um sources are nearly always spatially coincident with an \ion{H}{2} region ($>92\%$ of sources).  The spatial correlation with stellar associations identified in HST is also strong ($74\%$), and both of these correlations are significantly stronger than random association.
    \item The luminosity of the 21 \um correlates linearly with the attenuation-corrected H$\alpha$ emission over $\sim$ 3 orders of magnitude in luminosity (Figure \ref{fig:hacorr}).  However, there is a tail of sources with low 21 \um luminosity that are underluminous in H$\alpha$ emission. These may correspond to embedded sources.  However, potential embedded sources are comparatively rare and not uniformly linked to bright CO emission as would be expected if they were young massive clusters.
    \item While the 21 \um sources are strongly spatially associated with \ion{H}{2} regions, they are also associated with molecular clouds, as would be expected for these regions forming inside GMCs in a standard model of star formation.
\end{enumerate}
As a ``first-order'' approximation, the 21 \um compact sources seen by JWST are \ion{H}{2} regions that are already optically visible. However, only 14\% of the \ion{H}{2} regions have an associated 21 \um source.  The 21 \um sources are preferentially associated with the bright \ion{H}{2} regions and the low fraction of association seems to arise because the PHANGS--MUSE catalog of \ion{H}{2} regions is more sensitive than these initial 21 \um catalogs. This work demonstrates the capability to use broad band mid-IR imaging to efficiently identify the full set of compact star forming regions in a galaxy. 

\section*{Acknowledgments}

This research was conducted as part of the PHANGS collaboration. 

This work is based on observations made with the NASA/ESA/CSA JWST and Hubble Space Telescopes. All data were obtained from the Mikulski Archive for Space Telescopes (MAST) at the Space Telescope Science Institute, which is operated by the Association of Universities for Research in Astronomy, Inc., under NASA contract NAS 5-03127 for JWST and NASA contract NAS 5-26555 for HST. The JWST observations are associated with program 2107, and those from HST with program 15454. The JWST data can be accessed via
\dataset[10.17909/9bdf-jn24]{http://dx.doi.org/10.17909/9bdf-jn24}.

This paper makes use of the following ALMA data: \linebreak
ADS/JAO.ALMA\#2017.1.00886.L, \linebreak 
ADS/JAO.ALMA\#2018.1.01651.S. \linebreak 
ALMA is a partnership of ESO (representing its member states), NSF (USA) and NINS (Japan), together with NRC (Canada), MOST and ASIAA (Taiwan), and KASI (Republic of Korea), in cooperation with the Republic of Chile. The Joint ALMA Observatory is operated by ESO, AUI/NRAO and NAOJ.

Based on observations collected at the European Southern Observatory under ESO programmes 094.C-0623 (PI: Kreckel), 095.C-0473,  098.C-0484 (PI: Blanc), 1100.B-0651 (PHANGS-MUSE; PI: Schinnerer), as well as 094.B-0321 (MAGNUM; PI: Marconi), 099.B-0242, 0100.B-0116, 098.B-0551 (MAD; PI: Carollo) and 097.B-0640 (TIMER; PI: Gadotti).

HH and ER acknowledge the support of the Natural Sciences and Engineering Research Council of Canada (NSERC), funding reference number RGPIN-2022-03499.
MC gratefully acknowledges funding from the DFG through an Emmy Noether Research Group (grant number CH2137/1-1).
COOL Research DAO is a Decentralized Autonomous Organization supporting research in astrophysics aimed at uncovering our cosmic origins.
JMDK gratefully acknowledges funding from the European Research Council (ERC) under the European Union's Horizon 2020 research and innovation programme via the ERC Starting Grant MUSTANG (grant agreement number 714907).
TGW acknowledges funding from the European Research Council (ERC) under the European Union’s Horizon 2020 research and innovation programme (grant agreement No. 694343).
EJW, RSK, SCOG acknowledge funding provided by the Deutsche Forschungsgemeinschaft (DFG, German Research Foundation) -- Project-ID 138713538 -- SFB 881 (``The Milky Way System'', subprojects A1, B1, B2, B8, and P1).
MB acknowledges support from FONDECYT regular grant 1211000 and by the ANID BASAL project FB210003.
JK gratefully acknowledges funding from the Deutsche Forschungsgemeinschaft (DFG, German Research Foundation)  through the DFG Sachbeihilfe (grant number KR4801/2-1). 
KK gratefully acknowledges funding from the Deutsche Forschungsgemeinschaft (DFG, German Research Foundation) in the form of an Emmy Noether Research Group (grant number KR4598/2-1, PI Kreckel).
FB would like to acknowledge funding from the European Research Council (ERC) under the European Union’s Horizon 2020 research and innovation programme (grant agreement No.726384/Empire).
RSK acknowledges support from the European Research Council via the ERC Synergy Grant ``ECOGAL'' (project ID 855130), from the Heidelberg Cluster of Excellence (EXC 2181 - 390900948) ``STRUCTURES'', funded by the German Excellence Strategy, and from the German Ministry for Economic Affairs and Climate Action for funding in project ``MAINN'' (funding ID 50OO2206).
AKL gratefully acknowledges support by grants 1653300 and 2205628 from the National Science Foundation, by award JWST-GO-02107.009-A, and by a Humboldt Research Award from the Alexander von Humboldt Foundation.

%

\vspace{5mm}
\facilities{JWST(NIRCam, MIRI), HST(WFC3, ACS), ALMA, VLT(MUSE)}

\software{
astrodendro \citep{astrodendro},
astropy \citep{2013A&A...558A..33A,2018AJ....156..123A}, 
CIGALE \citep{cigale},
photutils \citep{photutils},
          }





\bibliography{ms,phangsjwst}{}
\bibliographystyle{aasjournal}


\suppressAffiliationsfalse
\allauthors
\end{document}